\documentclass[a4paper,9pt]{article}
\usepackage{pos}
\title{Lattice calculation of the \textit{R}-ratio smeared with Gaussian kernels}
\ShortTitle{}

\newcommand{\ket}[1]{\ensuremath{| {#1} \rangle }}
\newcommand{\bra}[1]{\ensuremath{\langle {#1} |}}

\author[1,2]{C. Alexandrou}
\author[2]{S. Bacchio}
\author*[3]{A. De Santis}
\emailAdd{alessandro.desantis@roma2.infn.it}
\author[4]{P. Dimopoulos}
\author[2]{J. Finkenrath}
\author[3]{R. Frezzotti}
\author[5]{G. Gagliardi}
\author[6]{M. Garofalo}
\author[1,2]{K. Hadjiyiannakou}
\author[7]{B. Kostrzewa}
\author[8]{K. Jansen}
\author[9]{V. Lubicz}
\author[6]{M. Petschlies}
\author[5]{F. Sanfilippo}
\author[5]{S. Simula}
\author[3]{N. Tantalo}
\author[6]{C. Urbach}
\author[10]{U. Wenger}

\affiliation[1]{Department of Physics, University of Cyprus, 20537 Nicosia, Cyprus}
\affiliation[2]{Computation-based Science and Technology Research Center, The Cyprus Institute, 20 Konstantinou Kavafi Street, 2121 Nicosia, Cyprus}
\affiliation[3]{Dipartimento di Fisica and INFN, Universit\`a di Roma Tor Vergata, Via della Ricerca Scientifica 1, I-00133 Roma, Italy}
\affiliation[4]{Dipartimento di Scienze Matematiche, Fisiche e Informatiche, Universit\`a di Parma and INFN, Gruppo Collegato di Parma, Parco Area delle Scienze 7/a (Campus), 43124 Parma, Italy}
\affiliation[5]{Istituto Nazionale di Fisica Nucleare, Sezione di Roma Tre, Via della Vasca Navale 84, I-00146 Rome, Italy}
\affiliation[6]{HISKP (Theory), Rheinische Friedrich-Wilhelms-Universit\"at Bonn,
	Nussallee 14-16, 53115 Bonn, Germany}
\affiliation[7]{High Performance Computing and Analytics Lab, Rheinische Friedrich-Wilhelms-Universit\"at Bonn, Friedrich-Hirzebruch-Allee 8, 53115 Bonn, Germany}
\affiliation[8]{NIC, DESY, Platanenallee 6, D-15738 Zeuthen, Germany}
\affiliation[9]{Dipartimento di Matematica e Fisica, Universit\`a Roma Tre and INFN, Sezione di Roma Tre, Via della Vasca Navale 84, I-00146 Rome, Italy}
\affiliation[10]{Institute for Theoretical Physics, Albert Einstein Center for Fundamental Physics,	University of Bern, Sidlerstrasse 5, CH-3012 Bern, Switzerland}

\abstract{The ratio $R(E)$ of the cross-sections for $e^+e^-\to$ hadrons and $e^+e^-\to \mu^+\mu^-$ is a valuable energy-dependent probe of the hadronic sector of the Standard Model. Moreover, the experimental measurements of $R(E)$ are the inputs of the dispersive calculations of the leading hadronic vacuum polarization contribution to the muon $g-2$ and these are in significant tension with direct lattice calculations and with the muon $g-2$ experiment. In this talk we discuss the results of our first-principles lattice study of $R(E)$. By using a recently proposed method for extracting smeared spectral densities from Euclidean lattice correlators, we have calculated $R(E)$ convoluted with Gaussian kernels of different widths $\sigma$ and central energies up to $2.5$~GeV. Our theoretical results have been compared with the KNT19~\cite{SMkeshavarzi2020g} compilation of experimental results smeared with the same Gaussian kernels and a tension (about three standard deviations) has been observed for $\sigma\sim 600$~MeV and central energies around the $\rho$-resonance peak.}

\FullConference{%
  The 39th International Symposium on Lattice Field Theory (Lattice2022),\\
  8-13 August, 2022 \\
  Bonn, Germany 
}

\begin{document}
\maketitle

\section{Introduction}

The $R$-ratio between the $e^+e^-$ cross-section into hadrons with that into muons has played a fundamental r\^ole in particle physics since its introduction in Ref.~\cite{Cabibbo:1970mh}. In recent years, the importance of the $R$-ratio has been mainly associated with the fact that the experimental results for $R(E)$ are the input of the dispersive calculations of the leading hadronic contribution (HVP) to the muon anomalous magnetic moment ($a_\mu$). The data-driven dispersive determinations of $a_\mu^\mathrm{HVP}$, reviewed in detail in Ref.~\cite{Aoyama:2020ynm}, are in strong tension (about four standard deviations) with the experimental measurement of $a_\mu$. On the other hand, lattice determinations of (partial) contributions to $a_\mu^\mathrm{HVP}$, obtained without any reference to the experimental measurements of $R$, are in much better agreement with the $a_\mu$ experiment~\cite{Borsanyi:2020mff}. 

In our recent work~\cite{Alexandrou:2022tyn} we performed a lattice QCD study of $R(E)$ and obtained first-principle results, discussed in this talk, that we then compared directly with $e^+e^-$ collision experiments with no reference to the muon $g-2$. More precisely, by using the method proposed in Ref.~\cite{Hansen:2019idp} and recently validated in Ref.~\cite{Bulava:2021fre}, we extracted the $R$-ratio smeared with normalized Gaussian kernels, $G_\sigma(\omega)=\exp(-\omega^2/2\sigma^2)/\sqrt{2\pi \sigma^2}$, according to
\begin{flalign}
	R_\sigma(E)=\int_{0}^\infty d\omega\, G_\sigma(E-\omega)\, R(\omega)\;. \label{eq:Rsigma}
\end{flalign}
We then compared our theoretical determinations of $R_\sigma(E)$ with experiments by smearing the $R(E)$ measurements with the same Gaussians. In this way, by varying $E$ and $\sigma$, we probed the $R$-ratio in Gaussian energy bins of different widths.  Around the $\rho$-resonance peak and at $\sigma\simeq 600$~MeV we managed to compute $R_\sigma(E)$ with an accuracy at the $1\%$ level. In this region our results are in tension by about three standard deviations with experiments.

\section{Methods}

In order to compute $R_\sigma(E)$, we relied on our effort within the Extended Twisted Mass Collaboration (ETMC) that produced a collection of state-of-the-art isospin symmetric QCD ensembles with four dynamical Twisted Mass quark flavours~\cite{Frezzotti:2000nk} at physical pion masses  (see Table~\ref{tab:ensembles} and Ref.~\cite{Alexandrou:2022amy}), together with the two-point Euclidean correlators of the quark electromagnetic current
\begin{flalign}
	V(t)=-\frac{1}{3}\sum_{i=1}^3\int d^3x \,\mathrm{T}\bra{0}J_i(x)J_i(0)\ket{0}\;.
\end{flalign}
In the previous formula $J_\mu=\sum_{f}q_f \bar \psi_f \gamma_\mu \psi_f$ with $f=\{u,d,s,c,b,t\}$, $q_{u,c,t}=2/3$ and $q_{d,s,b}=-1/3$. The connection between the correlators $V(t)$ and the $R$-ratio is given by the well known formula
\begin{flalign}
	V(t)=\frac{1}{12\pi^2}\int_{E_{th}}^\infty d\omega \, \omega^2 R(\omega)\, e^{-t\omega}\;,
	\label{eq:vtinfinite}
\end{flalign}
where the threshold energy $E_{th}$ is $2m_\pi$ in iso-symmetric QCD.
Theoretically $R(\omega)$ is a distribution, the spectral density of the correlator $V(t)$, and can conveniently be probed by using suitable smearing kernels. We considered Gaussian smearing kernels, providing a class of observables that are well localized in the energy domain, and computed $R_\sigma(E)$ on the lattice by using the method of Ref.~\cite{Hansen:2019idp}. The starting point of this approach is the following exact representation of the smearing kernel for $\omega>0$,
\begin{flalign}
	\frac{12\pi^2G_\sigma(E-\omega)}{\omega^2} =\sum_{\tau=1}^{\infty} g_\tau\, e^{-a\omega \tau}\;,
	\label{eq:DeltaRepExact}
\end{flalign}
where $\tau$ is an integer variable and $a$ is an arbitrary scale that, on the lattice, we identify with the lattice spacing. Once the coefficients $g_\tau\equiv g_\tau(E,\sigma)$ are known, $R_\sigma(E)$ can be computed according to
\begin{flalign}
	R_\sigma(E)=\sum_{\tau=1}^\infty g_\tau\, V(a\tau)\;.
	\label{eq:RRepExact}
\end{flalign}
Since the sums in Eqs.~(\ref{eq:DeltaRepExact}) and~(\ref{eq:RRepExact}) have necessarily to be truncated, the goal is to find a finite set of coefficients such that both the systematic and statistical errors on the resulting approximation to $R_\sigma(E)$ can be kept under control. 
As known, this numerical problem rapidly becomes ill-posed for $E>E_{th}$ and $\sigma\ll E$ (see Refs.~\cite{Hansen:2019idp,Bulava:2021fre} for illustrative numerical evidences of this fact).   The algorithm of Ref.~\cite{Hansen:2019idp} provides a regularization mechanism to the problem. We refer to Refs.~\cite{Hansen:2019idp,Bulava:2021fre} for extended discussions of this point.
In the method of Ref.~\cite{Hansen:2019idp} smearing kernels are represented as
\begin{flalign}
	K(\omega;\vec g) =\sum_{\tau=1}^{\tau_\mathrm{max}} g_\tau\, \left\{
	e^{-a\omega \tau}+e^{-a\omega(T-\tau)}
	\right\}\;,
	\label{eq:frepresentation}
\end{flalign}
where the second exponential takes into account the fact that on a lattice of finite temporal extension $T$, and with periodic boundary conditions in time, the correlator $V(a\tau)$ is given by
\begin{flalign}
	V(a\tau)=\frac{1}{12\pi^2}\int_{E_{th}}^\infty d\omega \, \omega^2 R(\omega)\, \left\{
	e^{-a\omega \tau}+e^{-a\omega(T-\tau)}\right\}\;.
	\label{eq:vtfinite}
\end{flalign}
In the implementation of the method described in full details in Ref.~\cite{Alexandrou:2022tyn}, the distance between the target kernel and its representation in terms of the coefficients $g_\tau$ has been measured by the functionals
\begin{flalign}
	A_\mathrm{n}[\vec g] = \int_{E_0}^\infty d\omega\, w_\mathrm{n}(\omega)\left\vert
	K(\omega;\vec g) - \frac{12\pi^2G_\sigma(E-\omega)}{\omega^2}
	\right\vert^2\;,
\end{flalign}
depending on the algorithmic parameter $E_0<E_{th}$ and corresponding to a class of weighted $L_2$-norms in functional space. We have considered the following weight functions
\begin{flalign}
	&w_\alpha(\omega) = e^{a\omega \alpha}\;,
	\qquad
	\alpha=\left\{0,\frac{1}{2},2^-\right\}\;,
	\qquad
	w_c(\omega)= \frac{1}{\sqrt{e^{a(\omega-E_0)}-1}},
	\label{eq:walpha}
\end{flalign}
that we distinguish by using the tag $\mathrm{n}=\{0,1/2,2^-,c\}$.
The regularization method adopted in Ref.~\cite{Hansen:2019idp} is the model-independent mechanism originally proposed by Backus and Gilbert~\cite{Backus}. The coefficients $\vec g$ are obtained by minimizing a linear combination,
\begin{flalign}
	W_\mathrm{n}[\vec g] =\frac{A_\mathrm{n}[\vec g]}{A_\mathrm{n}[\vec 0]} + \lambda\, B[\vec g]\;,
	\label{eq:wexpression}
\end{flalign}
of the norm-functional $A_\mathrm{n}[\vec g]$ and of the error-functional  
\begin{flalign}
	B[\vec g] =  
	B_\mathrm{norm} \sum_{\tau_1,\tau_2=1}^{\tau_\mathrm{max}} g_{\tau_1} g_{\tau_2}\, \mathrm{Cov}(\tau_1,\tau_2)\;,
	\label{eq:bdef}
\end{flalign}
where $\mathrm{Cov}(\tau_1,\tau_2)$ is the covariance matrix of the lattice correlator $V(a\tau)$. The relative normalization between the norm and error functionals has been set to
\begin{flalign}
	B_\mathrm{norm}=\frac{E^6}{\left(V(a \tau_\mathrm{norm})\right)^2}.
	\label{eq:relative_normalization}
\end{flalign}
At fixed values of the algorithmic parameters $\vec p=(\mathrm{n},\lambda, E_0,\tau_\mathrm{max},\tau_\mathrm{norm})$, the linear minimization problem 	
\begin{flalign}
\left.\frac{\partial W_\mathrm{n}[\vec g]}{\partial g_\tau}\right\vert_{\vec g=\vec g^{\vec p}}=0	
\end{flalign}
provides the coefficients $\vec g^{\vec p}$ and the corresponding approximation of $R_\sigma(E)$ according to
\begin{flalign}
	R_\sigma(E;\vec g^{\vec p}) = \sum_{\tau=1}^{\tau_\mathrm{max}} g_\tau^{\vec p}\, V(a\tau)\;.
\end{flalign}
The behaviour of $R_\sigma(E;\vec{g}^{\vec{p}})$ as function of the algorithmic parameters $\vec{p}$ and the procedure adopted to select our best estimates for $R_\sigma(E)$ are briefly discussed in Section~\ref{sec:dataanalysis} (see Ref.~\cite{Alexandrou:2022tyn} for more details).

\section{Materials}
\begin{table}{!t}
\begin{center}
\begin{tabular}{lcccc}
\textrm{ID}&
$L^3\times T$&
$a$ \textrm{fm}&
$aL$ \textrm{fm}&
$m_\pi$ \textrm{GeV}\\ \hline
\textrm{B64} & $64^3\cdot 128$ & 0.07961(13) & 5.09 & 0.1352(2) \\
\textrm{B96} & $96^3\cdot 192$ & 0.07961(13) & 7.64 & 0.1352(2) \\
\textrm{C80} & $80^3\cdot 160$ & 0.06821(12) & 5.46 & 0.1349(3) \\
\textrm{D96} & $96^3\cdot 192$ & 0.05692(10) & 5.46 & 0.1351(3) \\
\end{tabular}
\caption{\label{tab:ensembles}%
ETMC gauge ensembles used in this work, see Ref.~\cite{Alexandrou:2022amy} for more details.	}
\end{center}
\end{table}
The lattice gauge ensembles used in this work, generated by the ETMC, are listed in Table~\ref{tab:ensembles} and detailed in Ref.~\cite{Alexandrou:2022amy}. In particular, in order to better estimate the systematics associated with continuum extrapolations, we used the same mixed-action setup as described in Ref.~\cite{Alexandrou:2022amy,Frezzotti:2004wz}  and analyzed both the so-called Twisted Mass (TM) and Osterwalder-Seiler (OS) lattice regularized correlators $V(t)$. The results for $R_\sigma(E)$ obtained in the two regularizations differ by $O(a^2)$  cutoff effects~\cite{Frezzotti:2003ni,Frezzotti:2005gi} and must coincide within errors in the continuum. As customary, we considered separately the contributions corresponding to connected (C) and disconnected (D) fermionic Wick contractions to $V(t)$ and, in the case of the connected ones, also the contributions coming from the different flavours. We use e.g. the notation $R_\sigma^{ss,C,\mathrm{TM}}(E)$ for the ``strange-strange connected'' contribution to $R_\sigma(E)$ obtained from the correlator $V(t)$ in which the electromagnetic currents, in the Twisted Mass regularization, are both given by $-\bar s\gamma_\mu s/3$ and only fermionic connected Wick contractions are considered. The disconnected contribution, computed only in the OS regularization and including all flavours, will be denoted as $R_\sigma^{D}(E)$. 

In order to compare our theoretical results with experiments, we relied on the KNT19 compilation~\cite{keshavarzi2020g} of experimental results for the $R$-ratio, providing $R^\mathrm{exp}(E)$ in the range $E\in[0.216,11.1985]$~GeV together with the full covariance matrix that takes into account the correlation between the different experiments.

\begin{figure}[t!]
	\includegraphics[width=\columnwidth]{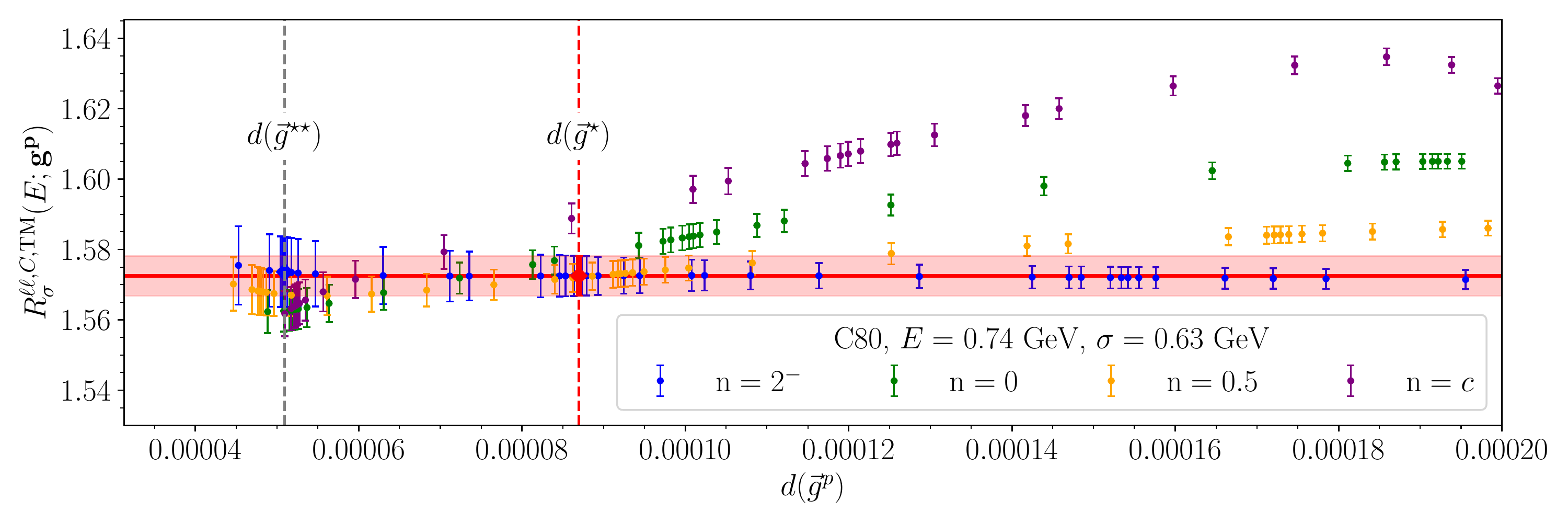}
	\caption{\label{fig:llcstability} \emph{Top-panel}: Example of the stability analysis procedure  in the case of the light-light connected contribution to $R_\sigma(E)$ for the TM regularization, on the C80 ensemble, at energy $E=0.74$ GeV and $\sigma=0.63$ GeV.  The errors on the points are statistical and the different colours refer to different weight functions, see Eq.~(\ref{eq:walpha}). The plot shows the variation of the results for different values of the reconstruction figure-of-merit $d(\vec{g}^{\vec{p}})$ defined in Eq.~(\ref{eq:smalld}). The two points $R_\sigma(E;\vec g^{\star})$ and $R_\sigma(E;\vec g^{\star\star})$, that we use respectively to pick the central value of $R_\sigma(E)$ and to estimate the residual systematic error (see Eq.~(\ref{eq:stars})), are marked by vertical dashed lines. The horizontal red band is the error $\bar{\Delta}_\sigma(E)$, calculated according to Eq.~(\ref{eq:deltabar}), that takes into account both the statistical and the systematic uncertainties. Different weight functions have a significant impact on the stability of the $R_\sigma(E)$ at varying $d(\vec{g}^p)$. The points obtained by setting $\mathrm{n}=2^-$ (blue points) are remarkably stable also at large values of $d(\vec{g}^p)$. }
\end{figure}

\section{Data analysis}
\label{sec:dataanalysis}
We considered three different values of $\sigma$,  namely  $
\sigma_1=0.44~\mathrm{GeV}$, $\sigma_2=0.53~\mathrm{GeV}$, $\
\sigma_3=0.63~\mathrm{GeV}$ and central energies $E$ in the range $[0.21,2.54]$~GeV. 
All our results have been obtained by fixing $E_0=0.21$~GeV and $\tau_\mathrm{max}=T/2+1$, corresponding respectively to $65$, $97$, $81$ and $97$ on the B64, B96, C80 and D96 ensembles.  The analysis described in the following has been performed for all the correlators, ensembles, regularizations, values of energy and $\sigma$.

On each gauge ensemble we set $\tau_\mathrm{norm}=1$ in the case of the connected contributions and $\tau_\mathrm{norm}=0$ in the case of the disconnected contributions, see Eq.~(\ref{eq:relative_normalization}).  In order to quantify the systematic error associated with the necessarily imperfect reconstruction of the smearing kernel, we studied $R_\sigma(E;\vec g^{\vec p})$ as a function of the normalized $L_2$--norm at $\mathrm{n}=0$ (also in the case where $\vec g^{\vec p}$ has been obtained with $\mathrm{n} \neq 0$), 
\begin{flalign}
d(\vec g^{\vec p})=\sqrt{\frac{A_0[\vec g^{\vec p}]}{A_0[\vec 0]}}\;.
\label{eq:smalld}
\end{flalign}
We quoted our best estimate for $R_\sigma(E)$ by selecting a result from the region of the statistically dominated regime (see Figure~\ref{fig:llcstability} for an example of this analysis), i.e. the region of small values of $d(\vec g^{\vec p})$ where the results are stable, within the statistical errors $\Delta_\sigma^\mathrm{stat}(E;\vec g^{\vec p})$. For large values of $d(\vec g^{\vec p})$ the results corresponding to the different weight functions and/or different values of $\lambda$ are substantially different, simply because the reconstructed kernels are very different from the target ones and among themselves. Conversely, for sufficiently small values of $d(\vec g^{\vec p})$ the results of $R_\sigma(E;\vec g^{\vec p})$ tend to agree within the statistical errors that in this regime grow because of the ill-posedness of the numerical problem. 

We selected our central-value estimates for $R_\sigma(E)$ and quantified the residual systematic error from the results for $R_\sigma(E;\vec g^{\vec p})$ corresponding to the conditions 
\begin{flalign}
\frac{A_{2^-}[\vec g^\star]}{A_{2^-}[\vec 0] }=10 B[\vec g^\star]\;,
\qquad
\frac{A_{2^-}[\vec g^{\star\star}]}{A_{2^-}[\vec 0]}= B[\vec g^{\star\star}]\;.
\label{eq:stars}
\end{flalign}
The central values of our results correspond to $R_\sigma(E)\equiv R_\sigma(E;\vec g^{\star})$ and the associated statistical errors  to $\Delta_\sigma^\mathrm{stat}(E)\equiv\Delta_\sigma^\mathrm{stat}(E;\vec g^{\star})$.
Our choice of the relative normalization of the functionals in Eq.~(\ref{eq:wexpression}) is such that $R_\sigma(E;\vec{g}^\star)$ and $R_\sigma(E;\vec{g}^{\star\star})$ are both inside the region of the statistically dominated regime in most of the cases. When this didn't happen the difference $R_\sigma(E;\vec g^{\star})-R_\sigma(E;\vec g^{\star\star})$ has been found to give a conservative estimate of the residual systematic uncertainty.
In particular, we introduced the quantity
\begin{flalign}
P_\sigma(E)=\frac{R_\sigma(E;\vec g^{\star})-R_\sigma(E;\vec g^{\star\star})}{\Delta_\sigma^\mathrm{stat}(E;\vec g^{\star\star})}
\label{eq:syspull}
\end{flalign}
as a measure of the statistical compatibility with zero of the difference between the results obtained at $\vec g^{\star}$ and $\vec g^{\star\star}$ and estimated the systematic error $\Delta^{\mathrm{rec}}_\sigma(E)$ due to the imperfect reconstruction of the target kernel according to
\begin{flalign}
\Delta^{\mathrm{rec}}_\sigma(E) = 
\left\vert R_\sigma(E;\vec g^{\star})-R_\sigma(E;\vec g^{\star\star}) \right\vert 
\mathrm{erf}\left(
\frac{\left\vert P_\sigma(E)\right\vert}{\sqrt{2}}
\right).
\label{eq:syserror}
\end{flalign}
The error $\bar{\Delta}_\sigma(E)$ resulting from the stability analysis procedure is thus given by 
\begin{flalign}
\bar{\Delta}_\sigma(E) = \sqrt{\left[\Delta^\mathrm{stat}_\sigma(E)\right]^2+\left[\Delta^\mathrm{rec}_\sigma(E)\right]^2}.
\label{eq:deltabar}
\end{flalign}

We performed a data-driven estimation of the systematic errors associated with finite-volume effects by using the results obtained on the B64 and B96 ensembles differing only for the number of lattice points and, therefore, for the physical volumes. In particular we considered the quantity
\begin{flalign}
P^{L}_\sigma(E)=
\frac{R_\sigma\left(E;\frac{3L}{2}\right) - R_\sigma(E;L)}{
	\sqrt{\left[\bar{\Delta}_\sigma\left(E;\frac{3L}{2}\right)\right]^2 + \left[\bar{\Delta}_\sigma(E;L)\right]^2}}\;,
\label{eq:volumepull}
\end{flalign}
where $\bar{\Delta}_\sigma(E;L)$ is the error on $R_\sigma(E;L)$ extracted from the stability analysis performed on the B64 ensemble ($aL\sim 5$~fm) while $\bar{\Delta}_\sigma(E;3L/2)$ and $R_\sigma(E;3L/2)$ are the corresponding quantities extracted from the B96 ensemble.  An estimate of the systematics associated with possible residual finite-volume effects has been obtained by considering
\begin{flalign}
&
\Delta^{L}_\sigma(E)=
\max_{\mathrm{reg}=\{\mathrm{OS},\mathrm{TM}\}}\Bigg\{
\left\vert R^\mathrm{reg}_\sigma\left(E;\frac{3L}{2}\right) - R^\mathrm{reg}_\sigma(E;L)\right\vert
\mathrm{erf}\left(
\frac{\left\vert P^{L,\mathrm{reg}}_\sigma(E) \right\vert}{\sqrt{2}}
\right)
\Bigg\}.
\label{eq:volumesys}
\end{flalign}
\begin{figure}[t!]
	\begin{center}
		\includegraphics[scale=0.4]{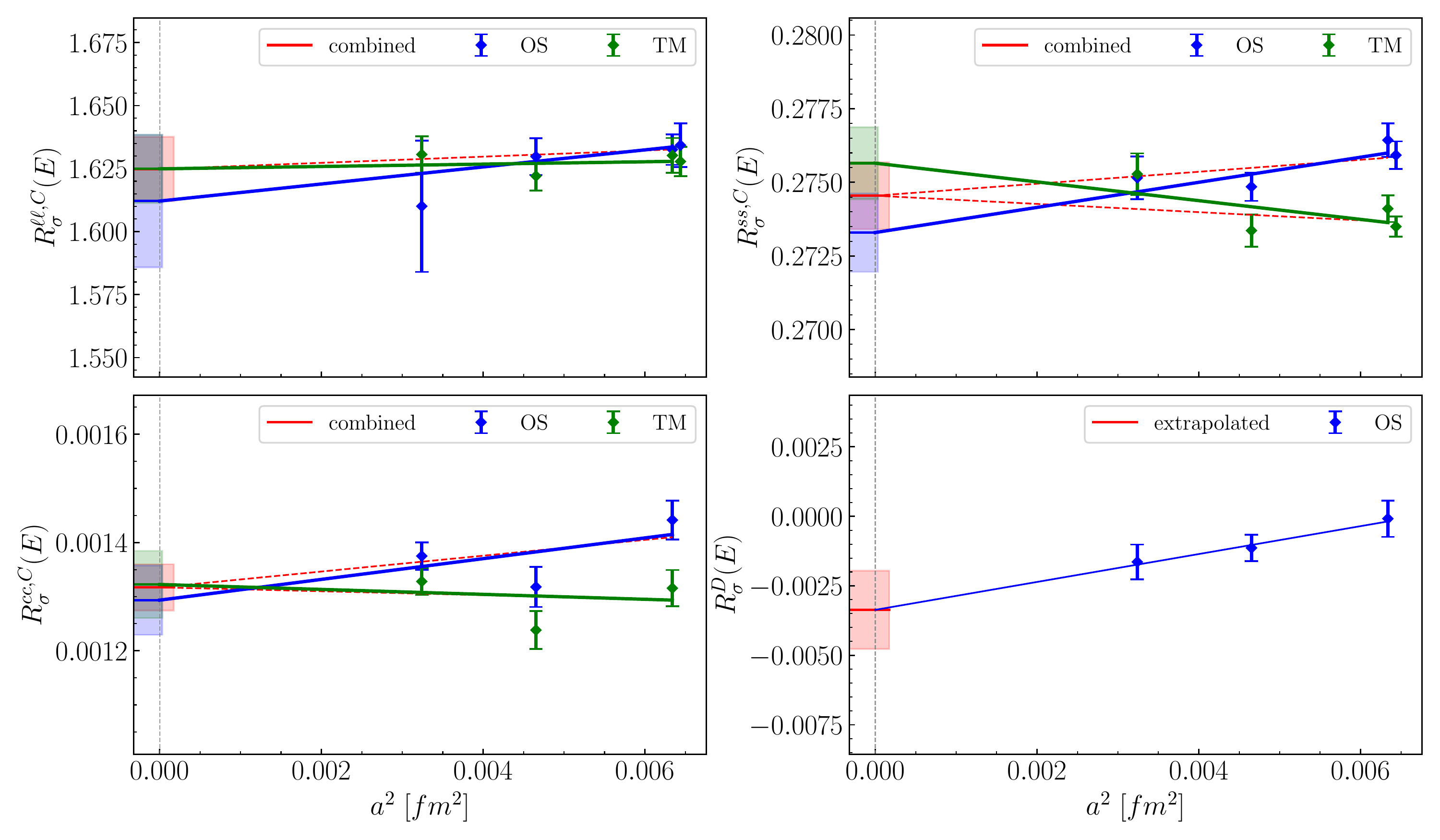}
		\caption{\label{fig:contmain} Continuum extrapolations of the different contributions to $R_\sigma(E)$ at $E=0.79$~GeV and $\sigma=0.63$~GeV. The blue and green points correspond respectively to the OS and TM lattice regularizations. In the case of the connected contributions we performed both correlated-constrained (red) and uncorrelated-unconstrained linear extrapolations in $a^2$ and found them to be compatible within errors in all cases. The disconnected contribution has been computed in the OS regularization only and extrapolated linearly in $a^2$. 
			In the case of $R^{\ell\ell,C}_\sigma(E)$ and $R^{ss,C}_\sigma(E)$ there are two points for each regularization at the coarsest lattice spacing (slightly displaced on the $x$-axis to help the eye) corresponding to the ensembles B64 and B96 and, therefore, to different volumes. No significant finite-volume effects have been observed for all considered values of $E$ and $\sigma$.
		}
	\end{center}
\end{figure}

We performed both constrained (for the connected contributions) and unconstrained continuum extrapolations of the OS and TM lattice data by considering fit functions of the form $f^\text{reg}(a)=A^\text{reg}+B^\text{reg}a^2$. In the constrained extrapolations we fitted together the OS and TM data by performing a correlated $\chi^2$-minimization and by imposing $A^\text{TM}=A^\text{OS}=A$. An example of continuum extrapolations for the different contributions to $R_\sigma(E)$ at $E=0.79$~GeV and $\sigma=0.63$~GeV is shown in Figure~\ref{fig:contmain}.

To quantify the systematic error associated with our continuum extrapolations we studied the compatibility between constrained and unconstrained extrapolations. We considered the quantity
\begin{flalign}
P^{a,\mathrm{reg}}_\sigma(E)=\frac{A - A^\mathrm{reg}}{\sqrt{\left[\Delta A\right]^2 + \left[\Delta A^\mathrm{reg}\right]^2}},
\label{eq:apull}
\end{flalign}
where $A$ is the result of the combined extrapolation at the given values of $\sigma$ and $E$, $\Delta A$ its error while $A^\mathrm{reg}$ and $\Delta A^\mathrm{reg}$ are the results and errors of the unconstrained extrapolations.
We extracted the central values and errors of our continuum results from the constrained fits and estimated the systematic errors associated with the continuum extrapolations by considering
\begin{flalign}
\Delta^{a}_\sigma(E)=\max_{\mathrm{reg}=\{\mathrm{OS},\mathrm{TM}\}}\left\{
\left\vert A - A^\mathrm{reg}\right\vert
\mathrm{erf}\left(
\frac{\left\vert P^{a,\mathrm{reg}}_\sigma(E) \right\vert}{\sqrt{2}}
\right)
\right\}.
\label{eq:asys}
\end{flalign}

Our estimate of the total error, $\Delta_\sigma(E)$, has been finally obtained by summing in quadrature $\Delta_\sigma^\mathrm{stat}(E)$, $\Delta_\sigma^{\mathrm{rec}}(E)$, the errors associated with finite-volume effects ($\Delta_\sigma^{L}(E)$) and continuum extrapolations ($\Delta_\sigma^{a}(E)$). The quoted final result for $R_\sigma(E)$ is given by the sum of $R_\sigma^{\ell\ell}(E)$, $R_\sigma^{ss}(E)$, $R_\sigma^{cc}$ and $R_\sigma^{D}(E)$.

\section{Results}
\label{sec:results}	
%

%
%
\begin{figure}[t!]
	\includegraphics[width=\columnwidth]{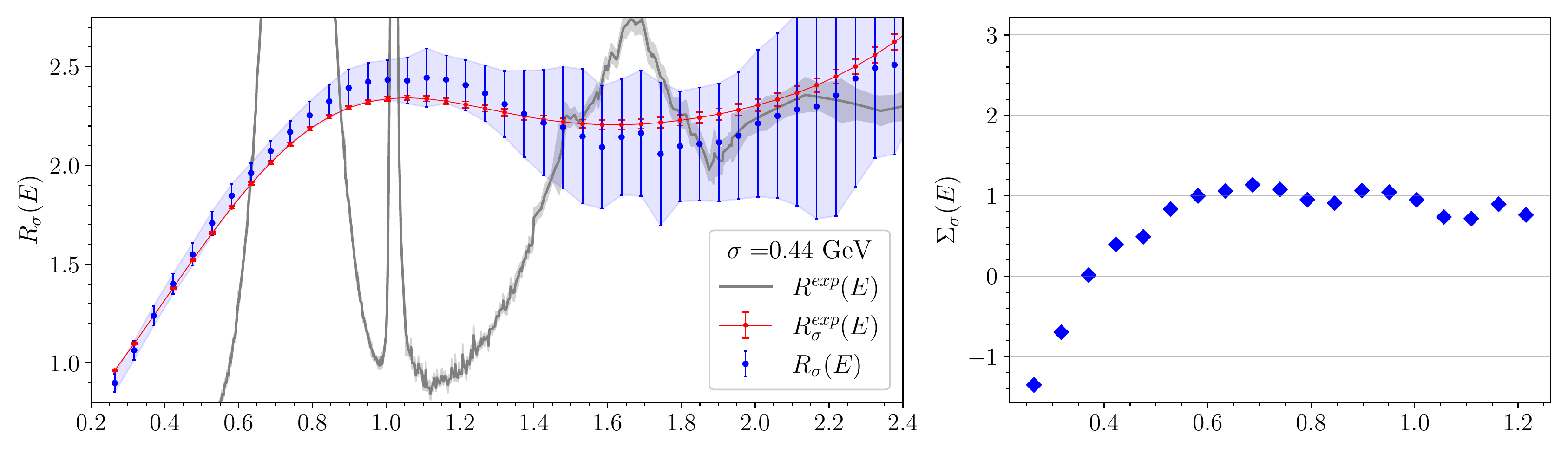}\\
	\includegraphics[width=\columnwidth]{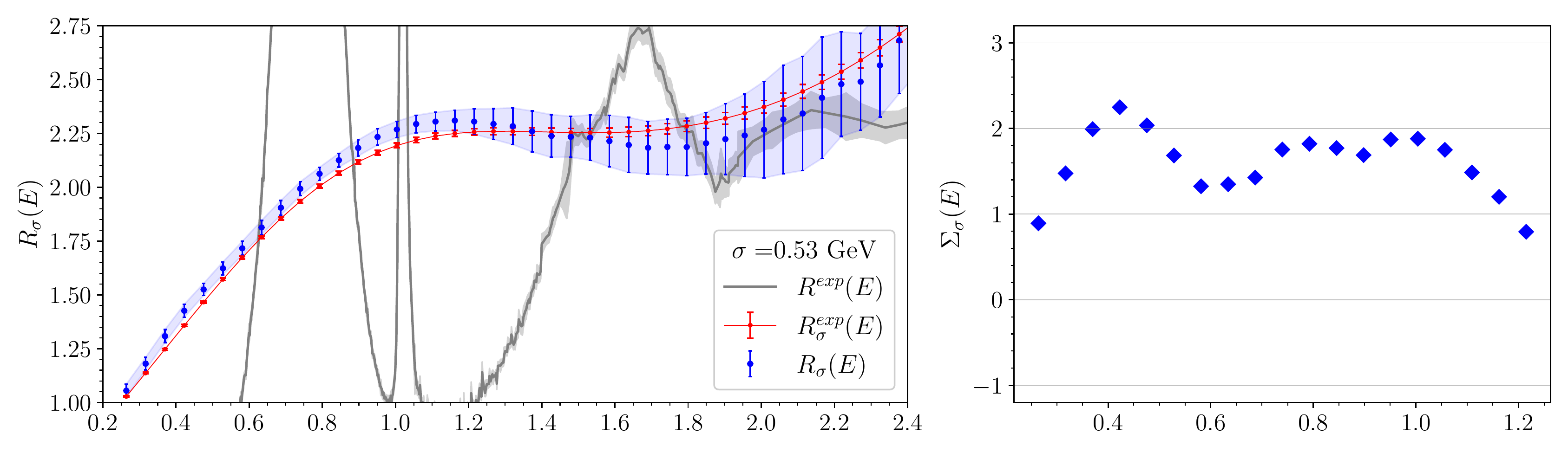}\\
	\includegraphics[width=\columnwidth]{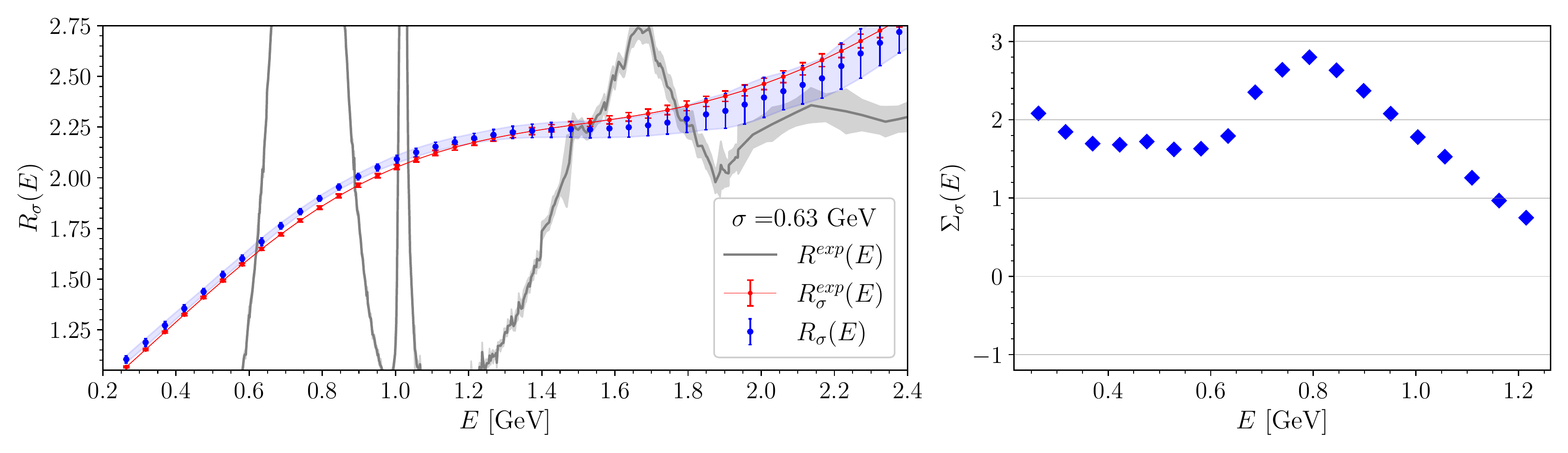}
	\caption{\label{fig:comparison} \emph{Left-plots}: Comparison of $R_\sigma(E)$ (blue points) and $R^\mathrm{exp}_\sigma(E)$ (red points) as functions of $E$ for $\sigma=0.44$~GeV (first row), $\sigma=0.53$~GeV (second row) and $\sigma=0.63$~GeV (third row) with $R_\sigma^\text{exp}(E)$. $R_\sigma^\text{exp}(E)$ has been obtained by generating 2000 bootstrap samples from the KNT19 compilation, each one simulating an independent measurement. Each sample has been integrated with the same Gaussian that defines $R_\sigma(E)$ and the final results for $R_\sigma^\text{exp}(E)$ has been obtained by taking the bootstrap average and standard deviation of the 2000 integrated samples.   \emph{Right-plots}: The quantity $\Sigma_\sigma(E)$ of Eq.~(\ref{eq:significance}) for $E<1.3$ GeV and for the three values of $\sigma$. Around $0.8$ GeV and for $\sigma=0.63$ GeV a tension of about three standard deviations is observed. The fact that the tension is smaller at smaller values of $\sigma$ is due to the increase of the theoretical error on $R_\sigma(E)$ since the reconstruction procedure becomes increasingly more difficult by lowering $\sigma$.}
\end{figure}

\begin{figure}[t!]
	\begin{center}
	\includegraphics[scale=0.4]{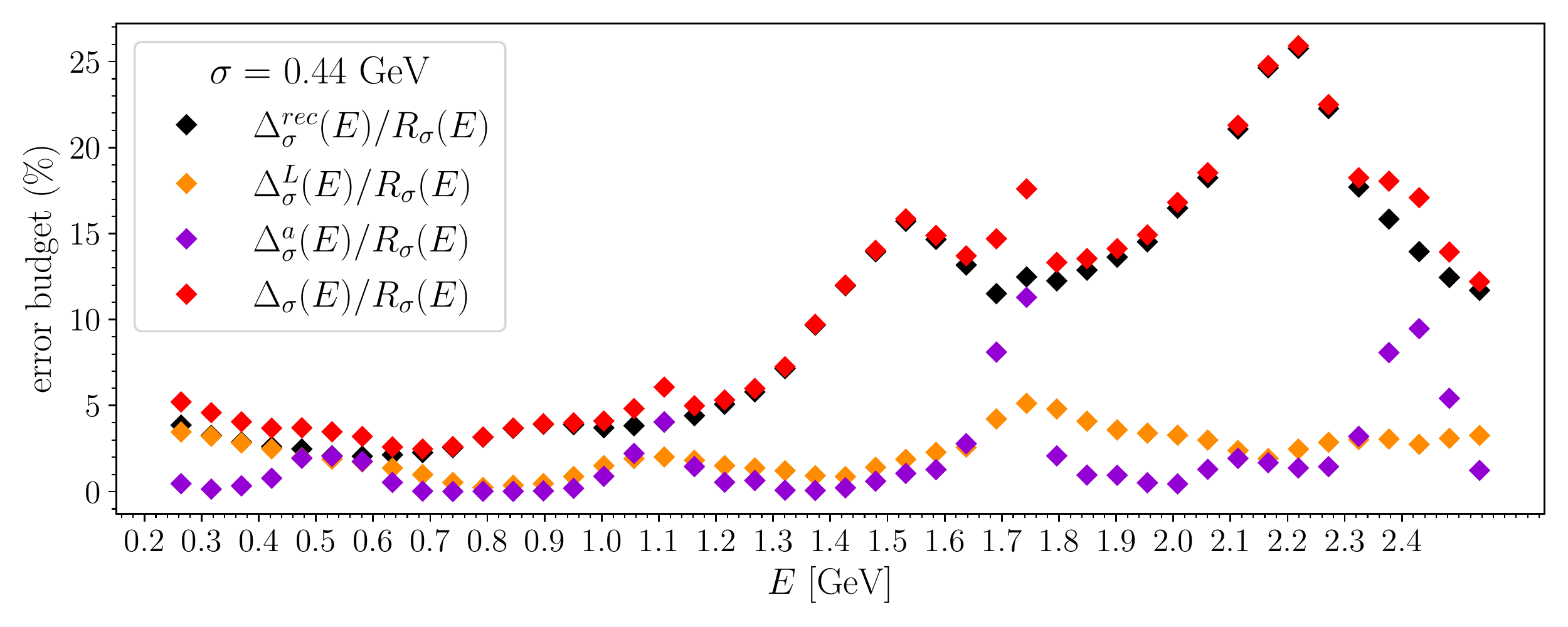}\\
	\includegraphics[scale=0.4]{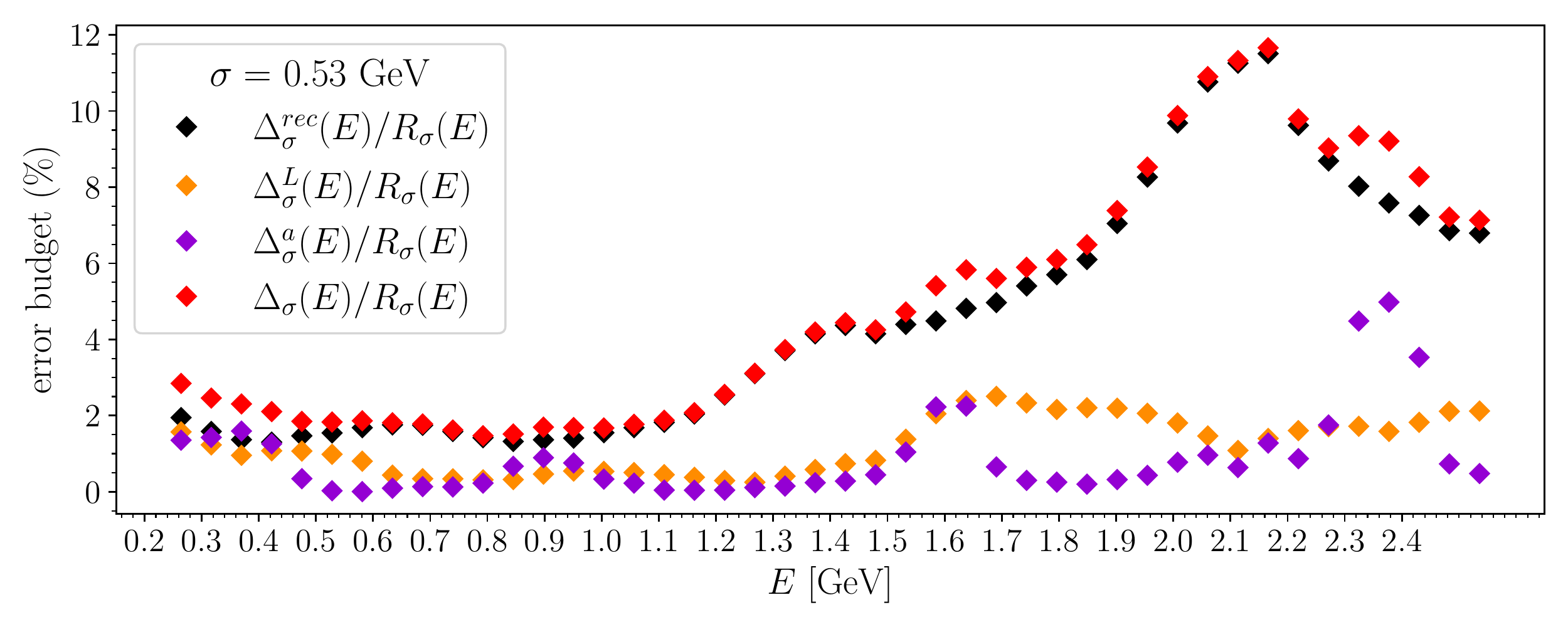}\\
	\includegraphics[scale=0.4]{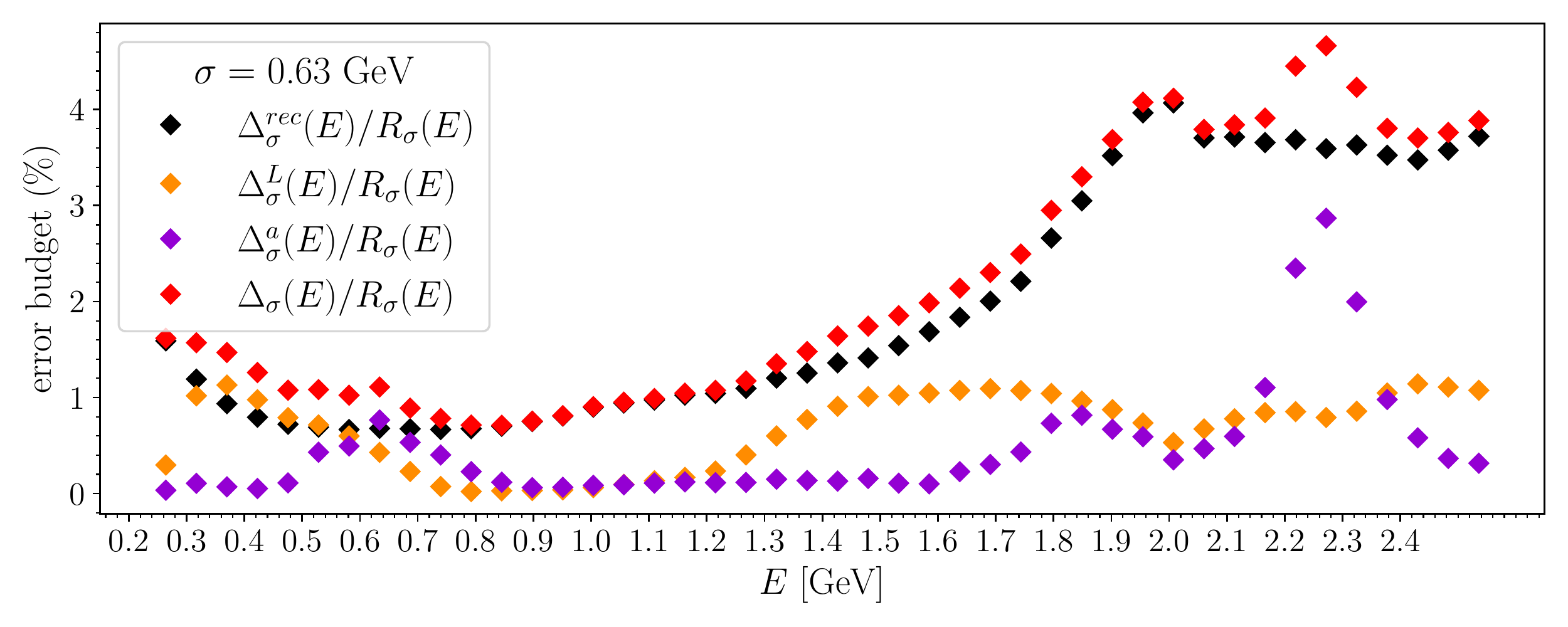}
	\caption{\label{fig:error_budget} Error budget for $R_\sigma(E)$ at $\sigma=0.44$~GeV (first row), $\sigma=0.53$~GeV (second row) and $\sigma=0.63$~GeV (third row). The red points correspond to the total relative error, $\Delta_\sigma(E)/R_\sigma(E)$. The black points are the statistical errors combined in quadrature with the systematics errors coming from the spectral reconstruction algorithm, $\bar \Delta_\sigma(E)$, divided by $R_\sigma(E)$. The violet and orange points are, respectively, our estimates of the relative systematics errors associated with the continuum extrapolations, $\Delta^a_\sigma(E)/R_\sigma(E)$, and finite volume effects, $\Delta^L_\sigma(E)/R_\sigma(E)$.}
\end{center}
\end{figure}

The comparison of our first-principles determination of $R_\sigma(E)$ with the experimental results $R_\sigma^{\text{exp}}(E)$ is shown in the left-plots of Figure~\ref{fig:comparison}. The plots show $R_\sigma(E)$ (blue points) and $R^\mathrm{exp}_\sigma(E)$ (red points) as functions of $E$ for $\sigma=0.44$~GeV (first row), $\sigma=0.53$~GeV (second row) and $\sigma=0.63$~GeV (third row). Our quoted final errors include the estimates of the different systematic uncertainties discussed in the previous section (see Figure~\ref{fig:error_budget} for the relative error budget). 

In order to properly interpret Figure~\ref{fig:comparison} it is very important to realize that the information contained into $R_\sigma(E)$ and $R_\sigma(E^\prime)$ for central energies such that $\vert E-E^\prime\vert \ll \sigma$ is essentially the same. Moreover, our theoretical results at different values of $E$ and $\sigma$ are obtained from the same correlators and, therefore, are correlated. It is also very important to stress that our lattice simulations have been calibrated by using hadron masses to fix the quark masses and the lattice spacing and, therefore, $R_\sigma(E)$ is a theoretical prediction obtained without using any input coming from $R^\mathrm{exp}_\sigma(E)$. In view of these observations, and of the fact that the extraction of spectral densities from Euclidean correlators is a challenging numerical problem, we consider the overall agreement between the theoretical and experimental data quite remarkable. 

Although our theoretical errors, $\Delta_\sigma(E)$, are still substantially larger than the experimental ones, $\Delta^\mathrm{exp}_\sigma(E)$, there is a tension between $R_\sigma(E)$ and $R^\mathrm{exp}_\sigma(E)$ in the region around the $\rho$ resonance. This can be better appreciated in the right panels of Figure~\ref{fig:comparison} where, for $E<1.3$~GeV, the plots show the ``pull'' 
\begin{flalign}
	\Sigma_\sigma(E) = \frac{R_\sigma(E)- R^\mathrm{exp}_\sigma(E)}{\sqrt{\left[\Delta_\sigma(E)\right]^2 + \left[\Delta^\mathrm{exp}_\sigma(E)\right]^2}}\;.
	\label{eq:significance}
\end{flalign}
Before ascribing this tension, of about three standard deviations, to new physics or to underestimated experimental uncertainties an important remark is in order. 

The calculation of $R_\sigma(E)$ that we have performed in Ref.~\cite{Alexandrou:2022tyn} is based an iso-symmetric $n_f=2+1+1$ lattice QCD calculation and, therefore, we have not calculated yet, from first principles, the contributions to $R_\sigma(E)$ coming from $b$-quarks and from the QED and strong isospin breaking corrections. Concerning the $b$-quark contribution, if sizeable, this would represent a positive correction to $R_\sigma(E)$ and thus, given the fact that $R^\mathrm{exp}_\sigma(E)$ is below $R_\sigma(E)$ in the region in which these are in tension, it can only lead to an enhancement of the observed discrepancy.

\begin{figure}[t!]
	\begin{center}
		\includegraphics[scale=0.4]{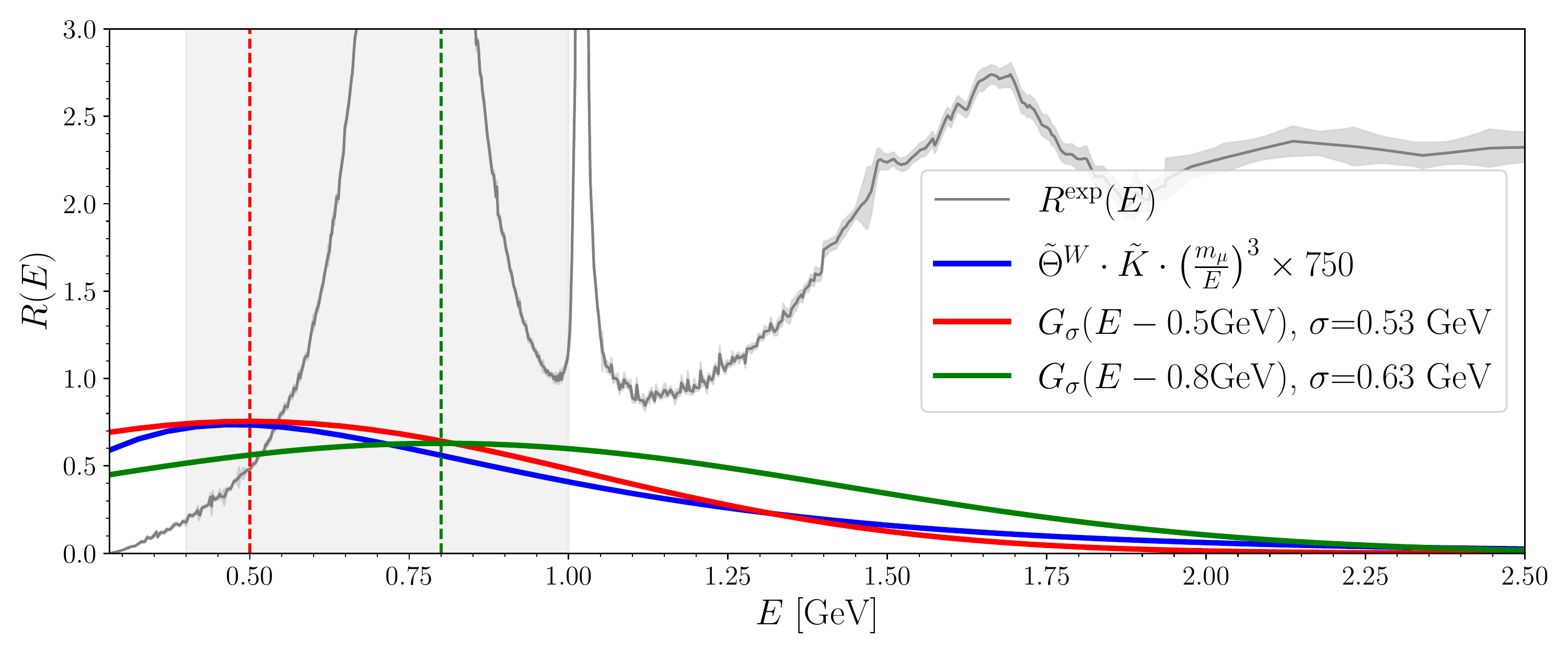}
		\caption{\label{fig:kernelW} 
		The Gaussian kernels with central energy 0.5 GeV and width 0.53 GeV (red) and central energy 0.8 GeV and width 0.63 GeV (green) are compared with the intermediate window kernel $\tilde{\Theta}^W\cdot \tilde{K}\cdot\left(\frac{E}{m_\mu}\right)^3$ (see e.g. Ref.~\cite{Alexandrou:2022amy} for the explicit expression). The red Gaussian is centred at the peak of the intermediate window kernel (vertical red line) that is shown in blue and normalized such that the heights of the two peaks coincide.  The green Gaussian is centred at the energy (vertical green line) where we observe the most significant tension (about $2.5$\% and $3$ standard deviations) between $R_\sigma(E)$ and $R_\sigma^\text{exp}(E)$. Using the red Gaussian we observe instead a 5\% tension corresponding to $2.2$ standard deviations, see Figure~\ref{fig:comparison}.}
	\end{center}
\end{figure}

Isospin breaking effects definitely have to be evaluated from first principles. 
Indeed, for very small values of $\sigma$ very large isospin breaking effects have to be expected at certain values of $E$, e.g. at very low energy where the channel $\pi^0+\gamma$ opens in QCD$+$QED and also close to other thresholds (see Refs.~\cite{Colangelo:2022prz,Hoferichter:2022iqe}).  
Nevertheless, we notice that in order to explain the observed tension at $E\sim 0.8$~GeV and $\sigma\sim 0.6$~GeV an isospin breaking effect larger than $2\%$ would be needed and this is hard to reconcile with the first principle lattice calculation performed in Ref.~\cite{Borsanyi:2020mff} of the isospin breaking corrections on closely related quantities, in particular on the intermediate window ($a_\mu^{\mathrm{HVP},W}$) contribution to $a_\mu^\mathrm{HVP}$. Indeed, the smearing kernel that in energy space defines $a_\mu^{\mathrm{HVP},W}$ is very similar in shape to the Gaussian kernel with central energy $E=0.5$~GeV and width $\sigma=0.53$~GeV  (see Figure~\ref{fig:kernelW}) and the isospin breaking effect on $a_\mu^{\mathrm{HVP},W}$ is found to be at the two permille level. We also note that, when $R(E)$ is convoluted with the quite different (but always very much spread out in energy)  kernels that define the long and short distance contributions to $a_\mu^\text{HVP}$ (see Ref.~\cite{Alexandrou:2022jlc}), the isospin breaking corrections w.r.t.\ iso-symmetric QCD remain very small, namely of about one permille~\cite{Borsanyi:2020mff} and three permille~\cite{HARLANDER2003244} respectively.

%

\section{
	\label{sec:conclusions}
	Conclusions
}
We presented the results of our recent non-perturbative theoretical study of the $e^+e^-$ cross-section into hadrons. We have calculated the $R$-ratio convoluted with Gaussian smearing kernels of widths between $440$~MeV and $630$~MeV and center energies up to $2.5$~GeV. We compared our first-principles theoretical results with the corresponding quantity obtained by using the KNT19 compilation~\cite{keshavarzi2020g} of $R$-ratio experimental data courteously provided by the authors.

In the region around the $\rho$ resonance we observe a tension between our theoretical determination and the experimental one, of about three standard deviations at $E\sim 800$~MeV and $\sigma\sim 600$~MeV. While the origin of this tension can be partly attributed to QED and strong isospin breaking corrections, we have  remarked that an isospin breaking corrections larger than $2\%$ would be required to fully reconcile our lattice data with experiments and that such a large correction is hardly conceivable in view of the few permille effects found in the related intermediate window contribution to $a_\mu^\mathrm{HVP}$ in ref.~\cite{Borsanyi:2020mff}.

A solid evidence of a significant tension between theory and experiment already emerged also from the comparison of the lattice calculations~\cite{Borsanyi:2020mff,Alexandrou:2022amy,FermilabLattice:2022smb,Ce:2022kxy} of the (window) contributions to $a_\mu^\mathrm{HVP}$ and the corresponding dispersive determinations~\cite{keshavarzi2020g}. Our results corroborate this evidence and, being totally unrelated to the muon $g-2$ experiment, highlight the fact that the tension is between experimental measurements of the $e^+e^-$ inclusive hadronic cross-section at energies around the $\rho$ resonance and first-principles Standard Model theoretical calculations. A phenomenological puzzle that certainly deserves further attention in the future.

\section*{Acknowledgments}
We warmly thank A.~Keshavarzi, D.~Nomura and T.~Teubner, the authors of the KNT19 combination~\cite{keshavarzi2020g} of $R$-ratio experimental measurements, for kindly providing us their results. We thank all members of ETMC for the most enjoyable collaboration. 
\bibliographystyle{JHEP}
\bibliography{bibli}

\end{document}